\documentclass[pdflatex,sn-nature]{sn-jnl}

\usepackage{graphicx}
\usepackage{amsmath,amssymb,amsfonts}
\usepackage{upgreek}    
\usepackage{bm}         
\usepackage{xcolor}     
\usepackage[normalem]{ulem}  
\usepackage{hyperref}
\usepackage{geometry}
\renewcommand{\rho}{\bm{\uprho}}

\begin{document}

\title[Refractive-index tomography of opaque tissue from backscattered light]
{Refractive-index tomography of opaque tissue from its own backscattered light}

\author[1]{\fnm{Tran Dinh} \sur{Hoang}}\equalcont{These authors contributed equally to this work.}
\author[1]{\fnm{Jaecheol} \sur{Cho}}\equalcont{These authors contributed equally to this work.}
\author[1]{\fnm{Thi Van Anh} \sur{Nguyen}}
\author[1]{\fnm{Eunyoung} \sur{Seong}}
\author[4]{\fnm{Joowon} \sur{Lim}}
\author[1]{\fnm{Jin Hee} \sur{Hong}}
\author[1]{\fnm{Yongwoo} \sur{Kwon}}
\author[2]{\fnm{Jun Wan} \sur{Kim}}
\author[2]{\fnm{Juhee} \sur{Yang}}
\author[3]{\fnm{Seokchan} \sur{Yoon}}
\author*[1]{\fnm{Sungsam} \sur{Kang}}\email{kssam01@gmail.com}
\author*[1]{\fnm{Wonshik} \sur{Choi}}\email{wonshik@korea.ac.kr}

\affil[1]{\orgdiv{Department of Physics}, \orgname{Korea University}, \orgaddress{\city{Seoul}, \postcode{02855}, \country{Republic of Korea}}}
\affil[2]{\orgdiv{Applied Electromagnetic Wave Research Center}, \orgname{Korea Electrotechnology Research Institute (KERI)}, \orgaddress{\city{Ansan}, \postcode{15588}, \country{Republic of Korea}}}
\affil[3]{\orgdiv{School of Biomedical Convergence Engineering}, \orgname{Pusan National University}, \orgaddress{\city{Yangsan}, \postcode{50612}, \country{Republic of Korea}}}
\affil[4]{\orgdiv{Optics Laboratory}, \orgname{\'Ecole Polytechnique F\'ed\'erale de Lausanne}, \orgaddress{\city{Lausanne}, \postcode{CH-1015}, \country{Switzerland}}}

\abstract{The refractive index (RI) is an intrinsic, label-free marker of a
living cell's dry mass and subcellular morphology, and hence of its
physiological state\cite{wang2011tissueRI,zangle2014mass}. Its three-dimensional (3D)
reconstruction has become a powerful
way to study cells and tissues in their native state, spanning cell
growth, drug response and disease diagnosis\cite{cooper2013chondrocyte,lee2024organoid,hugonnet2025cancer}.
Yet this capability rests on a fundamental
constraint: the RI can be recovered only from light
transmitted through the specimen\cite{wolf1969fdt,choi2007tpm,kim2014whitelight,kamilov2016bpm,park2018qpi,lim2019hifi,leeHugonnet2022mbs}, which demands optical access to both
sides. The cells that matter most---those within thick tissues, intact
organs and living animals---are therefore out of reach. A tissue,
however, can illuminate its own cells from behind: light backscattered by
intrinsic tissue structures beneath a cell carries the same transmission information a
microscope would collect from the far side. Here we develop a
divide-and-conquer inverse-scattering framework that recovers this
transmission from the backscattering and reconstructs a cell's
3D RI. We demonstrate label-free, quantitative imaging of cells
within an engineered tissue, and a living
mouse through its intact skull, where we further quantify the dry mass
of individual osteocytes \emph{in vivo}. By removing the need for two-sided
access, this reflection-only approach extends RI tomography
into living tissue, enabling non-destructive, longitudinal imaging of
cells in their native environment.}


\maketitle

\section{Introduction}

Set by the local concentration of proteins, lipids and nucleic acids,
the refractive index (RI) is an intrinsic optical property that reports
a cell's dry mass and, through its spatial variations, the morphology
of subcellular compartments such as nuclei, nucleoli and lipid
droplets. Together, these readouts make the RI a direct marker of the pathological states of cells and
tissues\cite{wang2011tissueRI,zangle2014mass}.
Quantitative RI imaging has tracked single-cell dry-mass growth,
viability, the cell cycle, apoptosis and drug
response\cite{mir2011growth,cooper2013chondrocyte,liu2020qpi,polanco2022drug}.
The same label-free contrast resolves processes across diverse cell
types, from dendritic-spine remodeling in neurons\cite{cotte2013nano}
and immunological-synapse formation in CAR-T cells\cite{lee2020cart} to
red-blood-cell membrane dynamics in malaria and sickle-cell
disease\cite{park2008rbc,hosseini2016sickle}. Holographic flow cyto-tomography further scales RI imaging to
high-throughput measurement of single cells in suspension\cite{merola2017flow},
while at the multicellular scale RI tomography extends to spheroids and
organoids, enabling long-term, label-free monitoring of their growth,
internal dynamics, and drug
response\cite{yasuhiko2023clearing,lee2024organoid,moser2025rotational}. More recently, integrating RI imaging with deep learning has enabled
label-free cancer diagnostics, from automated classification of
cancerous tissue\cite{zhang2022slim} to virtual H\&E staining of
unlabeled cancer tissue directly from label-free phase
images\cite{hugonnet2025cancer}.

Over the past two decades, optical diffraction tomography (ODT) has
become the leading technique for reconstructing such three-dimensional (3D) RI
maps\cite{choi2007tpm,sung2009odt,kim2014whitelight,park2018qpi}, with
extensions to the dielectric tensor of anisotropic
tissues\cite{saba2021psodt,shin2022dtt,yeh2024pti}, chemical-bond
contrast\cite{zhao2022bondidt} and nonlinear
susceptibility\cite{hu2020harmonic}. ODT inverts the inhomogeneous
Helmholtz equation from measured scattered fields; under a
first-order (Born or Rytov) approximation\cite{wolf1969fdt} this yields
high-resolution RI maps of live
cells\cite{choi2007tpm,sung2009odt,kim2014whitelight}. For optically
thick samples these first-order approximations break down, prompting more
accurate forward models --- multi-slice beam-propagation methods (BPM), including their split-step non-paraxial variants, together with
recursive Born and Rytov solvers --- that account for multiple scattering at substantially
higher computational
cost\cite{tian2015ledarray,kamilov2016bpm,lim2019hifi,chowdhury2019intensity,chen2020multilayer,leeHugonnet2022mbs},
extending ODT beyond the weak-scattering regime to multicellular
samples.

A more fundamental limitation, however, remains. Because the RI of each
voxel is encoded in the phase accumulated by waves passing through it,
conventional ODT must record transmitted waves and therefore needs
optical access to \emph{both} sides of the specimen. This restricts it
to thin or excised samples and excludes most \emph{in vivo} settings,
intact animals and opaque industrial specimens. Several
reflection-geometry approaches circumvent this, but each carries its
own limitation. Placing a reflective substrate behind the sample
redirects transmitted waves back through it, converting reflection into
transmission\cite{li2025reflection}, but is incompatible with
\emph{in vivo} deep-tissue imaging. Exploiting diffuse multiple
scattering turns epi-illumination into an effectively transmissive
geometry\cite{ford2012obm,ledwig2021epi,abraham2023slidefree}, but only
works in the weak-scattering approximation. Epi-illumination
phase-gradient microscopy images opaque samples in reflection, but
recovers morphology and integrated phase rather than a quantitative 3D
RI map\cite{kandel2019epiglim}. Finally, time-gated
reflection-matrix methods\cite{yoon2020laser,kwon2023computational,kang2023multiscattering}
recover only a few discrete axial layers, while focal-shift
speckle-diffraction-tomography\cite{kang2023sdt} assumes a stratified
specimen. Because of these limitations, 3D RI imaging of cells inside
thick tissues, living animals and human patients --- where optical
access to the far side is fundamentally impossible --- remains an unmet
need.

Here we present an inverse-scattering framework that reconstructs the
3D RI of cells deep inside opaque tissue from the
tissue's own backscattering. Backscattered light traverses the cells of
interest on its round trip and therefore carries their transmission
information; however, recovering that transmission from the raw
backscattering is severely ill-posed. We first apply time-gated
detection to retain only the light that completes the round trip through
the cells, rejecting reflections from shallower depths. We then develop
a divide-and-conquer strategy that splits the problem into two
well-posed stages: fitting the backscattering with an axially sparse
stack of complex layers, and then sweeping one of these layers through
the volume of interest (VOI) while holding the others fixed. The two stages are combined to construct the angular
transmission fields of the VOI, to which a standard
BPM-based ODT algorithm is applied to reconstruct its 3D
RI. We demonstrate the framework on HT29 cells in a
collagen matrix, a standard
tissue-engineering model\cite{caliari2016hydrogel,hofer2021organoids}; and, most notably, on
osteocytes \emph{in vivo} within the intact skull of a living mouse.

\section{Results}

\pdfbookmark[2]{Working principle}{wp-bm}
\medskip\noindent\textbf{Working principle}\par
\begin{figure}[htbp]
\centering
\includegraphics[width=0.75\linewidth,keepaspectratio]{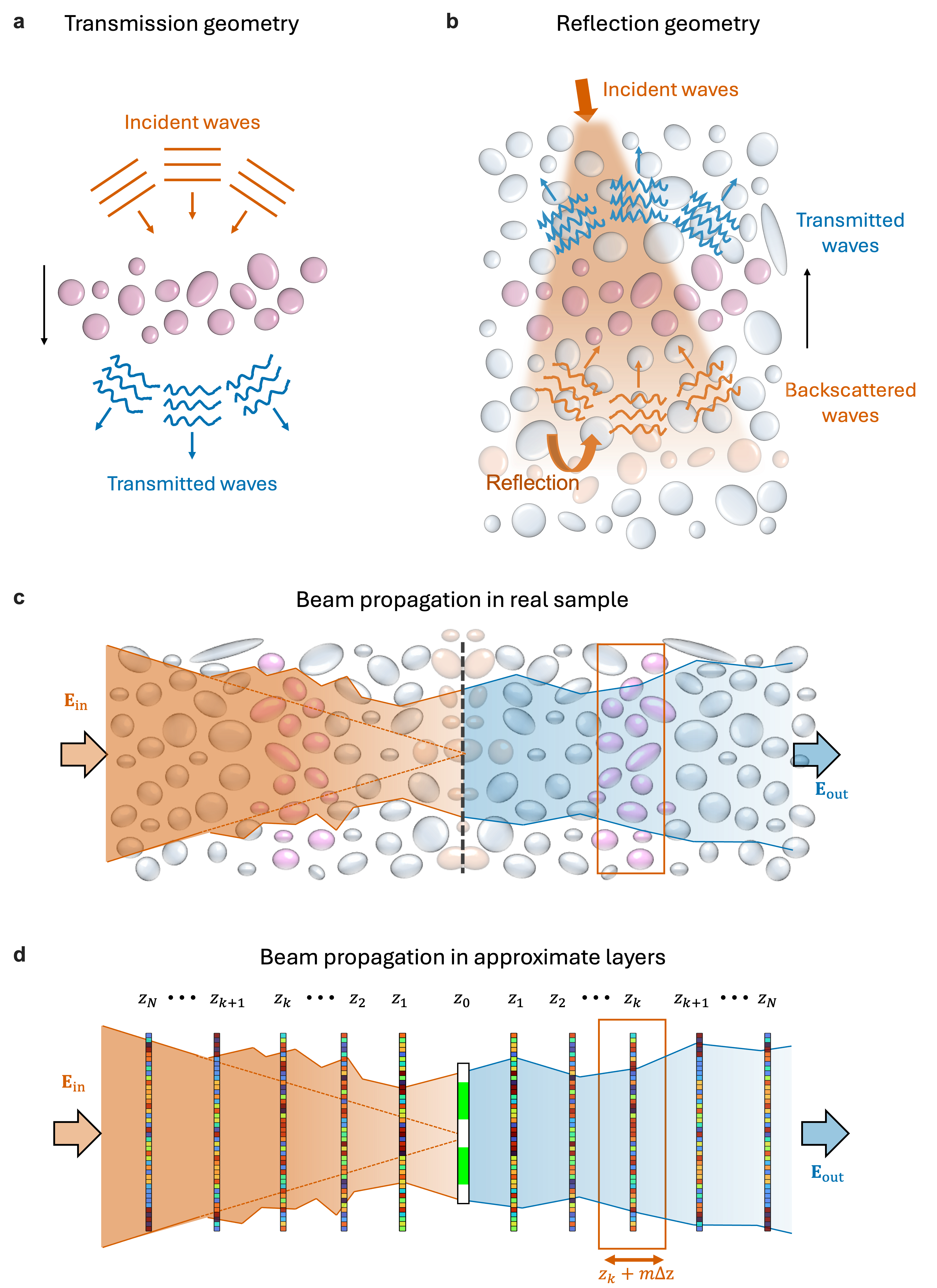}
\caption{\textbf{Concept of RI tomography from a scattering medium's own backscattering. a,}~Conventional transmission ODT
geometry, limited to thin transparent samples where access to both sides
is available. \textbf{b,} Proposed reflection-mode geometry. Incident
light penetrates the scattering medium and generates backscattered waves
from deeper structures (light beige) below the target volume (pink). These waves propagate upward, illuminate the target
volume from behind, and produce transmitted waves, which constitute the
experimentally measured backscattered signal.
\textbf{c,} Unfolding the light path. The round-trip reflection process
is reinterpreted as a double-pass
transmission geometry, placing the volume of interest (VOI, orange box)
in a transmission path. \textbf{d,} Multi-layer scattering
model. The scattering medium is approximated as a series of discrete
complex layers optimized to fit the experimentally measured
backscattered waves. The layer at axial position~\(z_{k}\), centered within the
VOI, is axially scanned (orange arrow) to axial positions \(z_{k}+m\Delta z\),
where \(m\) is an integer step index and \(\Delta z\) is the axial
step size, to retrieve the transmission fields of only the VOI, while the remaining
layers computationally compensate for scattering effects from the rest
of the medium.}
\label{fig:concept}
\end{figure}

\nobreak\smallskip\noindent
In conventional ODT, the
3D RI distribution of a specimen
is reconstructed by recording the waves transmitted through the
specimen for a set of illuminations (Fig.~\ref{fig:concept}a).
Unlike conventional ODT, our method operates in a reflection geometry
(Fig.~\ref{fig:concept}b). In this configuration, a portion of the incident light is
backscattered by structures within the scattering medium itself, located
deeper than the target volume of interest. These backscattered waves then propagate upward and transmit through
the VOI, effectively illuminating it from behind. However, multiple scattering in the scattering medium surrounding the VOI
distorts both waves---the back-illumination wave before it enters the VOI and
the transmitted wave after it exits---obscuring the true transmission
characteristics and preventing direct tomographic
reconstruction.

The most straightforward strategy to solve this problem is
to reinterpret the round-trip path as a double-pass transmission
(Fig.~\ref{fig:concept}c) and to model the entire optical path with a
densely voxelized, $\lambda/2$-spaced beam-propagation forward
model that is fitted to the measured backscattered waves. Several
prior reflection-mode tomographic approaches have followed this
route\cite{li2025reflection}. In practice, however, this
strategy suffers from two compounding limitations. First, a dense
model introduces one complex unknown per voxel along the full
round-trip path, so the inverse problem becomes severely
under-determined and prone to overfitting. Second, in the absence of
time gating, the recorded reflection contains a large
contribution from depths shallower than the VOI, which act as
unmodeled noise that further corrupts the fit. As a consequence,
existing reflection-mode reconstructions have been confined to thin
specimens or to configurations in which a strong, structured reflector
is deliberately placed behind the sample to dominate the
backscattered signal.

We address both limitations simultaneously. To isolate only the light
that completes the full round trip through the VOI to the reflecting
plane, we exploit \emph{time-gated} detection: the backscattered field
is recorded through a Mach--Zehnder interferometer driven by a
low-coherence source, so that only photons whose round-trip flight time
matches the coherence window set at the depth of the reflecting plane
contribute to the measurement, while reflections from shallower depths
and from depths beyond the reflecting plane are rejected (see Methods). On top of this gated measurement, we
replace the dense voxel grid with an \emph{axially sparse stack of
complex layers} (Fig.~\ref{fig:concept}d), each one fitted to
the backscattered waves. This sparse parameterization captures the
dominant multiple-scattering contributions of the
medium with orders of magnitude fewer unknowns than a dense BPM
model, suppressing overfitting while remaining expressive enough to
explain the gated data. By fitting this layered model to the
time-gated backscattering, we recover a set of complex amplitude
transmittance functions that best reproduce the observed signal.

The optimized complex amplitude transmittance of the layer at axial position~\(z_{k}\)~within the
VOI serves as a virtual, depth-resolved complex transmittance function of the VOI. By
scanning~\(z_{k}\)~in steps of~\(\Delta z\)~(typically $\lambda/2$, where $\lambda$ is the optical wavelength; see Methods) and repeating the
optimization, we obtain a set of depth-scanned complex transmittance functions, from
which we construct the angular transmission fields of the VOI and reconstruct 3D
RI maps using standard tomographic inversion algorithms. In essence,
splitting the inversion into two well-posed stages---a sparse-layer fit to
the backscattering data, then a depth scan of the layer in the VOI to recover its
transmission---circumvents the overfitting of a single full-voxel fit
over the entire round-trip volume.

\pdfbookmark[2]{Recovery of transmittance layers from backscattered light}{recovery-bm}
\medskip\noindent\textbf{Recovery of transmittance layers from backscattered light}\par
\begin{figure}[htbp]
\centering
\includegraphics[width=0.95\linewidth]{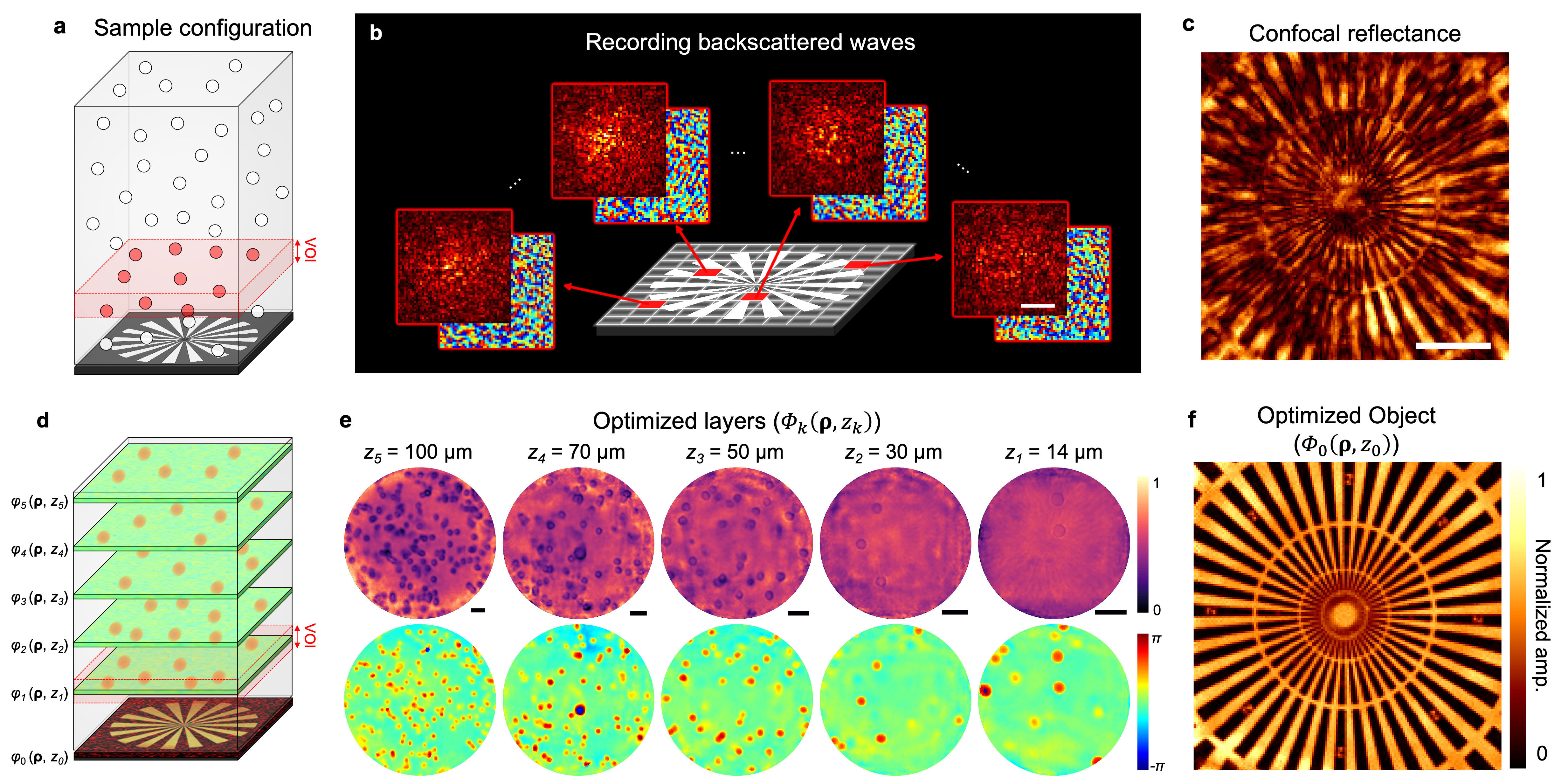}
\caption{\textbf{Multi-layer fitting to experimentally measured backscattered waves from a bead/Siemens-star phantom.}
\textbf{a}, Phantom sample: a thick scattering medium of randomly distributed 4.5-$\mu$m-diameter polystyrene beads with a Siemens-star reflecting layer placed underneath. The region marked in red denotes the volume of interest (VOI).
\textbf{b}, Data acquisition geometry. A focused illumination is raster-scanned across a 40~$\times$~40\,$\mu$m$^{2}$ area on the reflecting layer. The images show the recorded amplitude and phase maps for a few representative focus positions $\rho_{\text{in}} = (x_{\text{in}},y_{\text{in}})$ at the detection plane conjugate to the reflecting layer. Detection FOV per position: 20~$\times$~20\,$\mu$m$^{2}$.
\textbf{c}, Confocal reflectance image obtained by sampling the center pixel of the detection FOV in \textbf{b} at each scanned focus position $\rho_{\text{in}}$.
\textbf{d}, Multi-layer configuration. $\Phi_{1}(\rho,z_{1})$ is located within the VOI.
\textbf{e}, Optimized complex amplitude transmittance layers $\Phi_{k}(\rho,z_{k})$ for $k = 1\text{--}5$, retrieved at the model depths (amplitude, top; phase, bottom). The axial positions $z_{k}$ of the model layers (measured from the reflecting layer, $z_{0}=0$) are indicated above each panel.
\textbf{f}, Optimized reflecting layer $\Phi_{0}(\rho,z_{0})$ at the deepest model depth. Color bar, normalized amplitude. All scale bars in this figure are 10\,$\mu$m.}
\label{fig:experimental-validation}
\end{figure}

\nobreak\smallskip\noindent
We first validated the method on a phantom of known composition. The
phantom comprised 4.5-$\mu$m-diameter polystyrene beads
(RI = 1.595) randomly distributed in a background medium (RI = 1.56),
with a total thickness of 100 $\mu$m (Fig.~\ref{fig:experimental-validation}a). A reflective
Siemens-star pattern was placed beneath the scattering medium to serve as
a structured reflector. The VOI, whose 3D
RI map is to be reconstructed, is indicated by the red
shaded region in Fig.~\ref{fig:experimental-validation}a.

We measured the backscattered wavefields from this phantom using a
custom interferometric microscope that records time-gated
backscattered waves at a visible wavelength of $\lambda = 515$~nm
(see Methods for the detailed experimental setup). The illumination
was focused onto the reflecting layer and raster-scanned, and the
backscattered fields were recorded at each focal position
(Fig.~\ref{fig:experimental-validation}b). The numerical aperture (NA) of
the objective lens was 1.0, and the scan step was $\lambda/2$ over an
area of 40~$\times$~40\,$\mu$m$^{2}$.
The measured backscattered fields (Fig.~\ref{fig:experimental-validation}b)
were so severely distorted that the focused illumination dispersed
almost entirely into a speckle pattern, with the intensity at the focus
only 0.0055 times that of the surrounding time-gated scattered field.
In the resulting confocal reflectance image
(Fig.~\ref{fig:experimental-validation}c), the fine structures of the
Siemens-star pattern were almost completely obscured by strong
multiple scattering from the randomly dispersed beads.

We then fit our sparse-layer forward model to these measured
backscattered waves. The medium was approximated as six axially
discrete complex layers positioned at axial positions (measured from
the reflecting layer, \(z_{0}=0\))
\(\{ z_{0},z_{1},\ldots,z_{5}\} = \left\{ 0,\ 14,\ 30,\ 50,\ 70,\ 100 \right\}\)\,$\mu$m~(Fig.~\ref{fig:experimental-validation}d),
with each upper layer described by a complex amplitude transmittance function
\(\Phi_{k}(\rho,z_{k})\) for \(k = 1\text{--}5\), and the deepest
reflecting layer by an amplitude reflectance function \(\Phi_{0}(\rho,z_{0})\).
Here,~\(\rho = (x,y)\)~denotes the lateral coordinate on each layer.
The set of \(\{\Phi_{k}\}\) was optimized so that, at every focus
position, the backscattered wave predicted by this
layered model matched the corresponding measured wave shown in
Fig.~\ref{fig:experimental-validation}b (see Methods). The retrieved amplitude (top)
and phase (bottom) maps of the optimized transmittance layers are presented in
Fig.~\ref{fig:experimental-validation}e, and the optimized reflecting
layer is shown separately in Fig.~\ref{fig:experimental-validation}f.
The optimized model reproduced the measured
backscattered waves with a Pearson correlation of 0.50. The reflecting
layer~\(\Phi_{0}(\rho,z_{0})\) (Fig.~\ref{fig:experimental-validation}f) clearly reproduced the
Siemens-star pattern with a spatial resolution of $\lambda/2$. The transmittance layers visualized the phase retardation
of the beads near their respective depths, confirming the accuracy of
the optimization.

\pdfbookmark[2]{Reconstruction of angular transmission fields through the VOI}{angular-fields-bm}
\medskip\noindent\textbf{Reconstruction of angular transmission fields through the VOI}\par
\begin{figure}[htbp]
\centering
\includegraphics[width=0.95\linewidth]{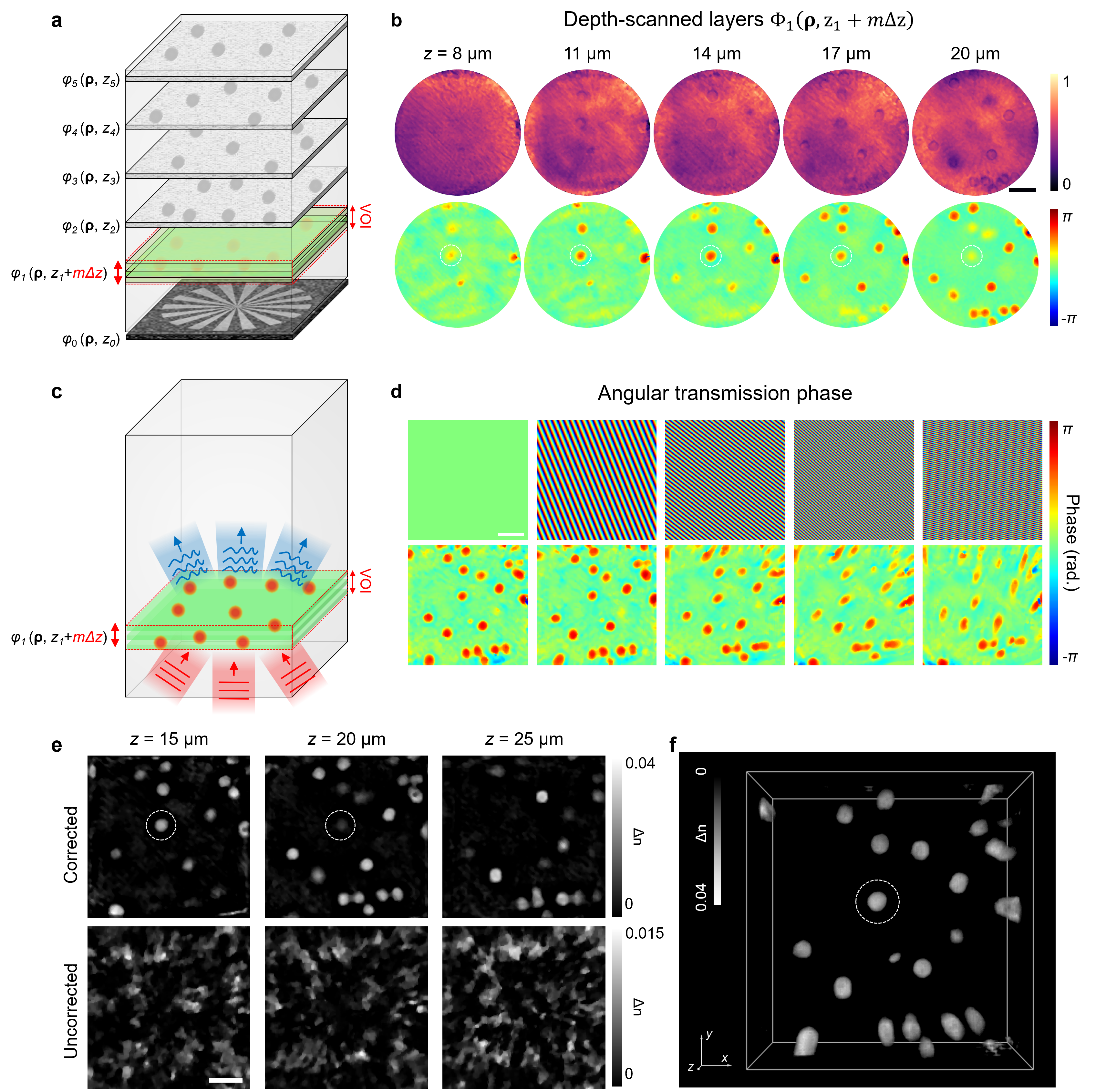}
\caption{\textbf{Reconstruction of the 3D RI map of the volume of interest (VOI) from the depth-scanned $\Phi_{1}$.}
\textbf{a}, Illustration of the axial scan of the transmittance layer $\Phi_{1}(\rho,z_{1})$ within the VOI through successive depths $z_{1} + m\Delta z$ ($m$ an integer).
\textbf{b}, Reconstructed $\Phi_{1}(\rho,z_{1}+m\Delta z)$ within the VOI for five representative depths $z = z_{1} + m\Delta z$ near $z_{1} = 14$~$\mu$m (amplitude, top; phase, bottom). The actual scan range of $z$ was 5--27\,$\mu$m. The bead marked by the white dashed circles is in focus near $z = 14$~$\mu$m and progressively defocuses at the other depths.
\textbf{c}, Illustration of the angular transmission fields through the VOI for different illumination angles.
\textbf{d}, Reconstructed angular transmission fields for selected incident wavevectors $(k_{x}^{\text{in}},k_{y}^{\text{in}})$. Top: phase maps of the incident planar waves. Bottom: phase maps retrieved from the optimized depth-scanned $\Phi_{1}$ in \textbf{b}, with the incident phase in the top row subtracted.
\textbf{e}, Depth section slices of the reconstructed RI tomogram with (top row, Corrected) the proposed scattering correction and without (bottom row, Uncorrected) scattering correction (see main text). Color bar, RI difference with respect to the background medium.
\textbf{f}, 3D-rendered RI map of the scattering-corrected tomographic reconstruction. In \textbf{e} and \textbf{f}, the dashed circles indicate the same bead as marked in \textbf{b}. All scale bars in this figure are 10\,$\mu$m.}
\label{fig:voi-reconstruction}
\end{figure}

\nobreak\smallskip\noindent
Building on the layered complex amplitude transmittance \(\{\Phi_{k}\}\) recovered
above, we next reconstructed the transmitted fields of the
VOI for various illumination angles. To this end, we axially scanned
the layer at depth~\(z_{1}\)~within the VOI in increments of $\Delta z = \lambda/2$ while keeping the
other layers fixed (Fig.~\ref{fig:voi-reconstruction}a). Here the VOI is the reconstruction target centered
on the scanned layer~\(z_{1}\), while the layers
\(z_{2},\dots,z_{5}\) model the scattering above the VOI. In principle, any of the layers~\(z_{k}\) could serve
as the scanned layer, each defining a VOI centered on its depth; we chose
\(z_{1}\) because it lies
closest to the reflecting plane~\(z_{0}\), where the spatial
resolution of the retrieved transmittance layer~\(\Phi_{k}\) is
highest, so that the VOI centered on it
attains the highest-resolution ODT reconstruction.

Fig.~\ref{fig:voi-reconstruction}b shows representative maps
of~\(\Phi_{1}(\rho,z_{1} + m\Delta z)\)~at several axial
positions \(z = z_{1} + m\Delta z\) (where \(m\) is an integer) near~\(z_{1} = 14\)\,$\mu$m; the actual
scan range of \(z_{1}\) was 5--27\,$\mu$m. As the layer depth was scanned, individual beads appeared only near their
respective depths; the bead marked by the white dashed circles, for instance, is
sharpest near~\(z = 14\)\,$\mu$m and blurs elsewhere. This confirms that the
depth-scanned~\(\Phi_{1}\)~maps serve as depth-resolved transmittance functions of the VOI.

In fact, each optimized transmittance layer \(\Phi_{1}\)~corresponds to
a synthetic aperture image or, equivalently, a confocal transmittance image with the focus set
at~\(z_{1} + m\Delta z\)~.
This can be understood from the wave-propagation model: the field
immediately after each layer is the product of the incident field and
the layer's complex amplitude transmittance at each point. Thus, the
optimized layer is given by the transmitted wave divided by the incident
wave, which is precisely the synthetic aperture image formed by
coherently combining many angled plane waves\cite{kim2011sam}. Meanwhile, multiple scattering from other depths is largely compensated by the~\(\Phi_{k}\)~of all the other layers (\(k \neq 1\)).

Using the
depth-scanned~\(\Phi_{1}(\rho,z_{1} + m\Delta z)\)~maps, we
next retrieved the transmitted
fields~\(E_{p}(\rho;\mathbf{q}_{\text{in}},z_{1})\) for
plane-wave illumination at various incident angles, set by the transverse
wavevector~\(\mathbf{q}_{\text{in}} = (k_{x}^{\text{in}},k_{y}^{\text{in}})\),~at
the focal depth~\(z_{1}\)~centered within the VOI~(Fig.~\ref{fig:voi-reconstruction}c).
Since each depth-scanned synthetic aperture image is a coherent
superposition of these angular transmission fields, we
recovered~\(E_{p}(\rho;\mathbf{q}_{\text{in}},z_{1})\)~by fitting them to
the set of~\(\Phi_{1}(\rho,z_{1} + m\Delta z)\)~(see Methods).

Fig.~\ref{fig:voi-reconstruction}d presents the retrieved phase
maps of~\(E_{p}(\rho;\mathbf{q}_{\text{in}},z_{1})\) for
representative \(\mathbf{q}_{\text{in}}\) (bottom row). The top row
of Fig.~\ref{fig:voi-reconstruction}d shows the phase maps of
the incident planar waves for each incidence angle, while the bottom
row shows the phase maps after subtracting these incident planar-wave
phases. The phase retardation induced by the spherical beads is clearly
visible. The bead images are also seen to elongate with the incidence angle.
More beads appear in Fig.~\ref{fig:voi-reconstruction}d than in any single depth section of Fig.~\ref{fig:voi-reconstruction}b, because the transmission fields integrate contributions from all depths within the VOI, projecting beads from every depth section onto a single image.

Finally, using the reconstructed angular transmission fields (Fig.~\ref{fig:voi-reconstruction}d),
we applied a BPM-based ODT algorithm
to reconstruct the RI map \(n\left( \mathbf{r} \right)\)
within the VOI\cite{kamilov2016bpm,lim2019hifi}. The algorithm models the medium as a stack of axial
slices of thickness $\lambda/2$, and \(n\left( \mathbf{r} \right)\) was
iteratively refined by minimizing the mean-squared difference between
the BPM-predicted transmission fields and the recovered
\(E_{p}\left( \rho;\mathbf{q}_{\text{in}},z_{1} \right)\) using the
adaptive moment estimation (ADAM) optimizer.

The impact of our scattering correction on the final RI tomogram is
illustrated in Fig.~\ref{fig:voi-reconstruction}e, where depth
section slices of the reconstructed RI map are shown for
both the scattering-corrected (top row) and uncorrected (bottom row)
cases. With scattering correction applied, the
4.5-$\mu$m-diameter beads embedded in the background medium were faithfully
recovered. The measured RI of the beads from the reconstructed
tomography was $1.591 \pm 0.005$, in close agreement with the expected
value of 1.595 against a background of 1.56. In contrast, when the same
BPM-based ODT reconstruction was applied directly to the
uncorrected angular transmission fields derived from the raw reflection
data, the algorithm failed
to reconstruct any of the embedded scatterers. A 3D-rendered visualization of the
scattering-corrected reconstruction is presented in
Fig.~\ref{fig:voi-reconstruction}f.

\pdfbookmark[2]{Refractive-index tomography of cells in a collagen matrix from intrinsic backscattering}{collagen-bm}
\medskip\noindent\textbf{Refractive-index tomography of cells in a collagen matrix from intrinsic backscattering}\label{sec:cells-in-collagen-only}\par
\begin{figure}[htbp]
\centering
\includegraphics[width=\linewidth]{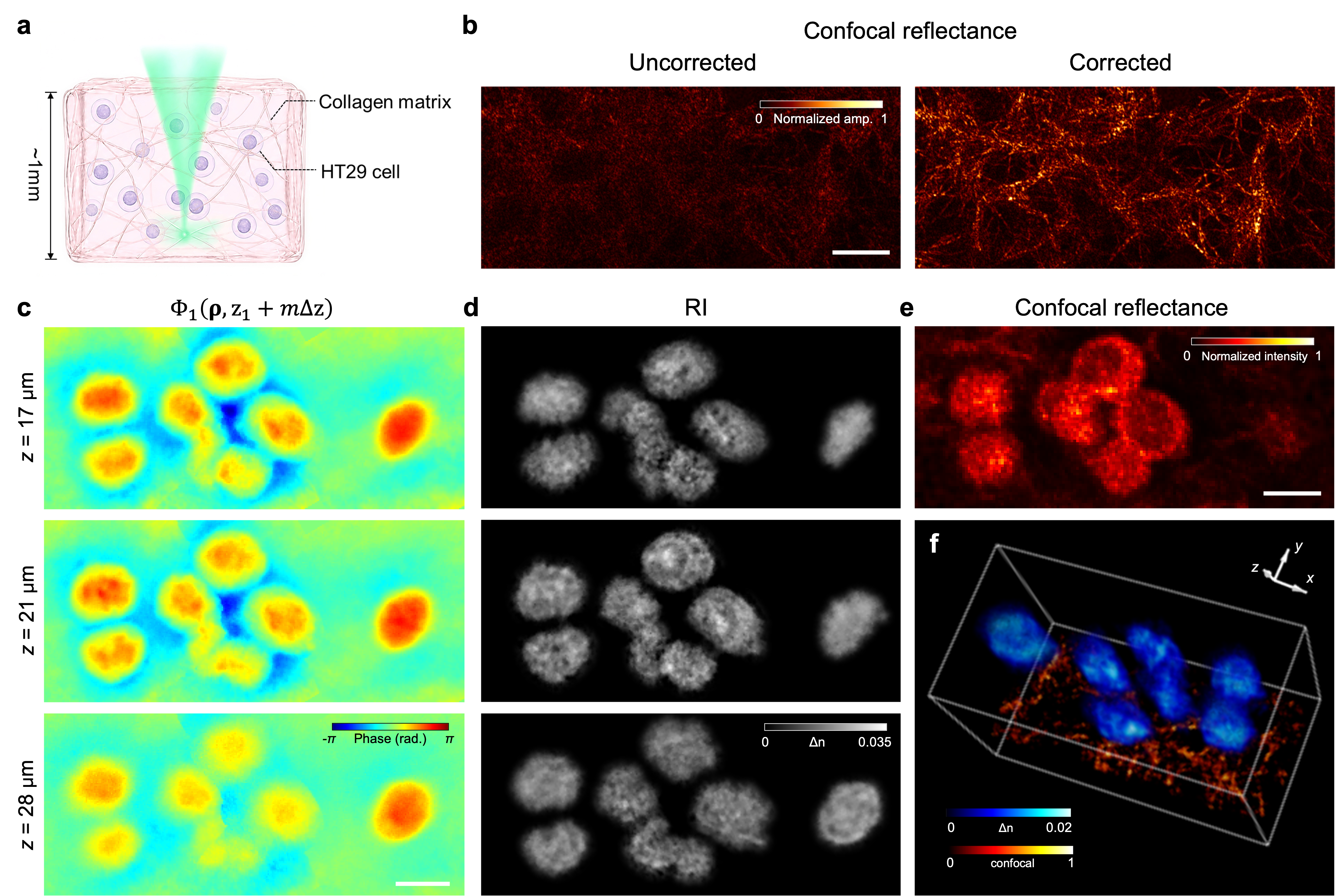}
\caption{\textbf{Refractive-index tomography of cells in a collagen matrix from intrinsic backscattering.}
\textbf{a,} Sample geometry: cells embedded in a collagen matrix, with the objective focused on the underlying collagen so that the backscattered light originates from the collagen itself.
\textbf{b,} Confocal reflectance image of the collagen layer at the focal plane without (left, Uncorrected) and with (right, Corrected) multi-layer scattering correction; scattering from the overlying cells and matrix is removed and fine collagen features are resolved.
\textbf{c,} Depth-scanned reconstructions of the optimized transmittance layer \(\Phi_{1}(\rho, z_{1} + m\Delta z)\) at three representative depths, each labeled by its axial position $z = z_{1} + m\Delta z$ ($z_{0}=0$).
\textbf{d,} Corresponding cross-sectional RI tomograms at the same depths.
\textbf{e,} Confocal reflectance image of the cells at $z = 21~\mu$m (corresponding to \textbf{c}, \textbf{d}).
\textbf{f,} 3D rendering of the reconstructed cell RI tomogram (blue) overlaid with the scattering-corrected confocal reflectance image (hot colormap). Color bars: RI difference $\Delta n$ relative to the background medium (blue) and normalized confocal reflectance intensity (hot). All scale bars in this figure are 10\,$\mu$m.}
\label{fig:cells-in-collagen-only}
\end{figure}

\nobreak\smallskip\noindent
Cell-laden collagen hydrogels are a canonical platform for tissue
engineering and 3D \emph{in vitro} biology. To demonstrate
RI tomography of cells in this geometry, we embedded HT29
colorectal cancer cells in a collagen matrix (see Methods). With the
objective focused on the collagen fibers,
we recorded the backscattered signal and
reconstructed the 3D RI map of the cells above the focal plane
(Fig.~\ref{fig:cells-in-collagen-only}a). In the raw confocal reflectance image
(Fig.~\ref{fig:cells-in-collagen-only}b, left), scattering from the
overlying cells and matrix obscures the collagen fiber structure at the
focal plane; multi-layer optimization compensates for both contributions and
restores the fiber structure from the intrinsic
backscattering of the in-focus collagen fibers alone
(Fig.~\ref{fig:cells-in-collagen-only}b, right).

Axially scanning the optimized transmittance layer
\(\Phi_{1}(\rho, z_{1} + m\Delta z)\) yielded depth-resolved amplitude
and phase maps of the cell distribution across the volume of interest
(Fig.~\ref{fig:cells-in-collagen-only}c), and tomographic inversion of
these layers produced sharp cross-sections that resolved individual
cells with intracellular detail at three representative depths
(Fig.~\ref{fig:cells-in-collagen-only}d). Nuclei and nucleoli invisible
in the individual phase maps emerge clearly in the reconstructed
sections, and the tomogram section at \(z = 21~\mu\)m matches the
confocal reflectance image acquired at the same depth
(Fig.~\ref{fig:cells-in-collagen-only}e), confirming the fidelity of
the reconstruction. The RI values of the cell nuclei, nucleoli, and cell
body were measured to be 1.348--1.352, 1.359--1.361, and 1.352--1.357,
respectively, against a background collagen medium RI of 1.339, falling
within a range comparable to values measured in transmission
geometry~\cite{sung2009odt,kim2014whitelight}. A 3D rendering maps
their volumetric distribution throughout the VOI
(Fig.~\ref{fig:cells-in-collagen-only}f).

Notably, reconstructions obtained from the weak intrinsic backscattering
of the collagen matrix were consistent with those of the same sample
placed on a flat glass reflector, confirming that
the sample's own reflection suffices in place of an external mirror.

These results demonstrate quantitative 3D RI tomography
of cells in engineered tissue from intrinsic backscattering alone. Because it does not rely on transmitted
light, the method is insensitive to sample thickness, enabling non-destructive, label-free monitoring
of cell distribution, dry mass, and morphology in cell-laden hydrogel
constructs for tissue
engineering, drug screening, and disease modeling\cite{caliari2016hydrogel}.

\pdfbookmark[2]{In vivo RI tomography of osteocytes within an intact mouse skull}{skull-bm}
\medskip\noindent\textbf{\emph{In vivo} RI tomography of osteocytes within an intact mouse skull}\label{sec:through-skull}\par
\begin{figure}[htbp]
\centering
\includegraphics[width=\linewidth]{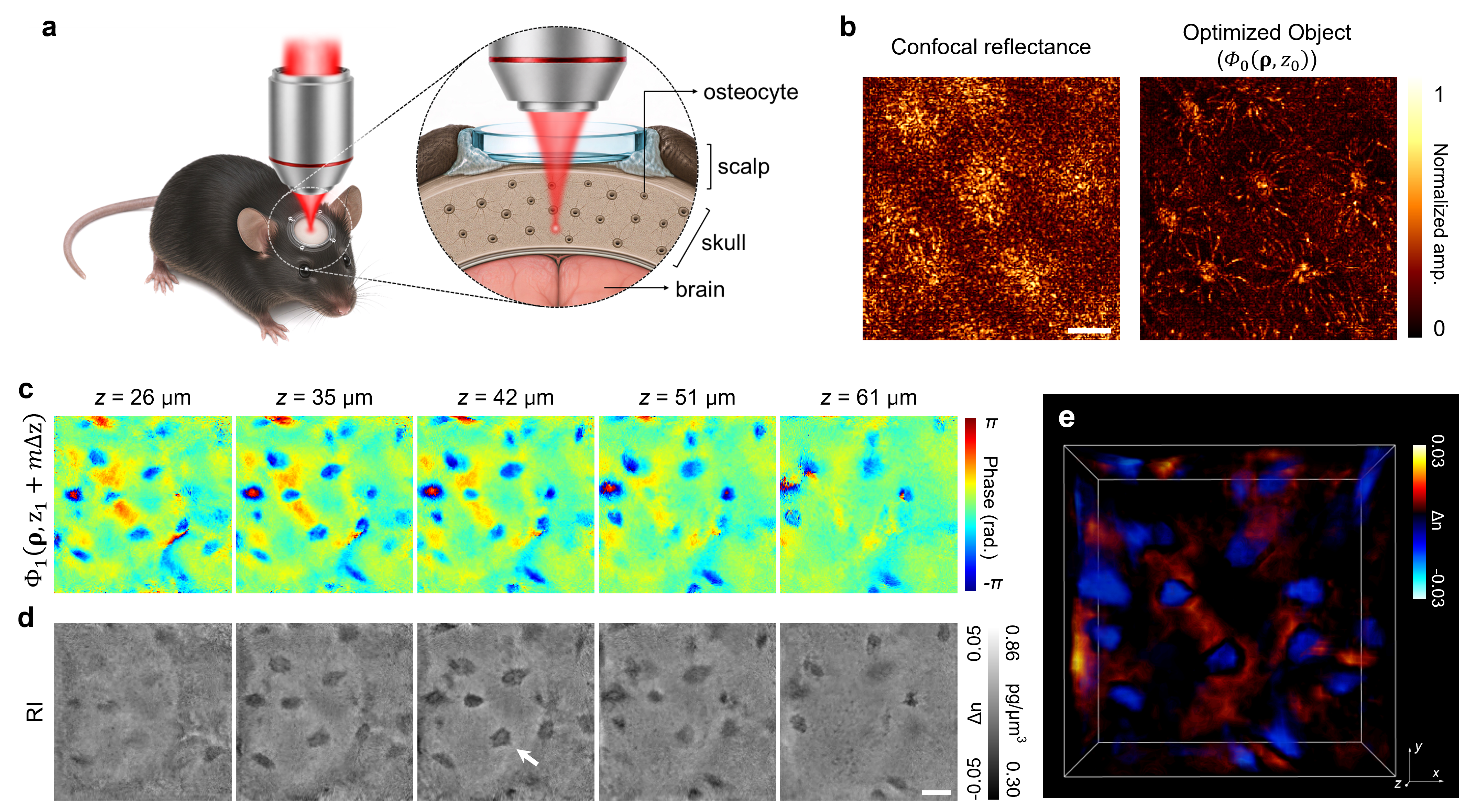}
\caption{\textbf{\emph{In vivo} RI tomography of osteocytes within an intact mouse skull.}
\textbf{a,} Imaging geometry: a head-fixed mouse imaged in epi-mode through an intact skull capped with a cover glass. The objective is focused inside the skull and the 3D RI map of the osteocytes above the focal plane is reconstructed, with the bone matrix in the skull serving as an intrinsic back-reflector.
\textbf{b,} Confocal reflectance image of the skull layer 140\,$\mu$m below the surface before scattering correction (left) and the optimized reflecting layer $\Phi_{0}(\rho,z_{0})$ after multi-layer fitting (right); the overlying-skull scattering is suppressed and fine structures such as osteocyte processes become resolved. Color bar, normalized amplitude.
\textbf{c,} Depth-scanned reconstructions of the optimized transmittance layer $\Phi_{1}(\rho,z_{1}+m\Delta z)$ at five representative depths above the reflecting layer $z_{0}$, each labeled by its axial position $z = z_{1} + m\Delta z$ ($z_{0}=0$). Color bar, phase in radians.
\textbf{d,} Corresponding cross-sectional RI tomograms at the same depths. The white arrow marks an example osteocyte whose dry mass is estimated to be 1348\,pg. Color bar, RI difference relative to the background ($\Delta n$) and dry-mass density.
\textbf{e,} 3D rendering of the reconstructed osteocyte RI tomogram. Color bar, RI difference $\Delta n$. All scale bars, 20\,$\mu$m.}
\label{fig:fig5}
\end{figure}

\nobreak\smallskip\noindent
We next applied the framework in a far more demanding setting: target
cells embedded in a living, strongly scattering tissue. Imaging entirely in epi-mode, we
reconstructed osteocytes inside the intact skull of a head-fixed mouse
(Fig.~\ref{fig:fig5}a). For a 20-week-old mouse with a skull thickness
of $\sim$200\,$\mu$m, we removed the scalp and attached a circular glass
coverslip to the exposed parietal bone using a biocompatible adhesive
(see Methods for details of sample preparation). For this measurement,
we acquired the data with a previously established 1.3\,$\mu$m system of
similar design to the 515\,nm setup\cite{kwon2023computational} (see
Methods). The objective was
focused 140\,$\mu$m beneath the skull surface, so that the bone matrix
at the focal plane acted as the intrinsic reflector. The refractive
index of the osteocytes lying above the focal plane was reconstructed,
with no exogenous label and no external mirror.

The difficulty of this regime is apparent in the raw confocal
reflectance image of the skull layer in the focal plane (Fig.~\ref{fig:fig5}b,
left), where scattering from the overlying skull obscures the
microscopic structures of the osteocytes. Multi-layer optimization separates this overlying
contribution from the true reflecting layer \(\Phi_{0}(\rho, z_{0})\),
whose recovered map clearly resolves the microscopic details of
individual osteocyte processes (Fig.~\ref{fig:fig5}b, right).

As in the collagen-matrix case, axially scanning the optimized
transmittance layer \(\Phi_{1}(\rho, z_{1} + m\Delta z)\) produced
depth-resolved phase maps that capture the osteocytes above the focal
plane (Fig.~\ref{fig:fig5}c; five representative depths above the
reflecting layer \(z_{0}\)). Refractive-index maps at the same depths are shown in
Fig.~\ref{fig:fig5}d. Notably, the osteocytes exhibit a
lower RI than the surrounding bone matrix, consistent with
the weaker phase retardation they impart in each optimized phase map
(Fig.~\ref{fig:fig5}c). The measured
RI difference of the osteocytes from the surrounding bone
matrix ranged from $-0.020$ to $-0.035$. In a separate experiment, we
measured the average RI of the bone matrix to be 1.425
(see Methods). Using this value, we determined the absolute
RI values of the RI tomography, from which we obtained the
dry-mass density in each voxel (see color scale in
Fig.~\ref{fig:fig5}d) and the dry mass of each individual
osteocyte (see Methods). For example, the cell indicated by the white
arrow in Fig.~\ref{fig:fig5}d had a dry mass of 1348\,pg. A 3D rendering of the reconstructed
tomogram resolves their spatial arrangement throughout the imaged
volume (Fig.~\ref{fig:fig5}e).

Together, these results establish that the framework can retrieve
3D RI information from cells embedded within
intact, strongly scattering bone --- a regime that transmission-mode
tomography cannot reach. Because
osteocytes orchestrate bone mechanosensing and remodeling,
this \emph{in vivo} demonstration opens a route to longitudinal,
label-free interrogation of bone biology at the single-cell level.
Single-cell dry mass has illuminated skeletal-cell biology --- for
example, linking growth-plate chondrocyte dry mass to skeletal
proportions\cite{cooper2013chondrocyte} --- but only in cultured or
isolated cells. Our approach brings this label-free readout into the
intact living animal, enabling dry-mass studies under more physiologically relevant conditions --- for
instance, tracking osteocyte maturation and remodeling in ageing and
disease.

\section{Discussion}

We have demonstrated quantitative 3D RI
tomography of cells inside intact biological tissue without optical
access to the far side of the specimen. The most direct strategy ---
recasting the round trip as a double-pass transmission and fitting a
single, densely voxelized beam-propagation model to the
backscattering --- is fundamentally ill-posed in thick tissue,
which has confined earlier reflection-mode tomography to thin specimens
or to samples backed by a deliberately dominant reflector. We overcome
this on two fronts. Time-gated detection isolates the backscattering
from the depth of interest and rejects out-of-gate light before
fitting. A divide-and-conquer inversion then replaces the monolithic
dense fit: the first stage fits only an axially sparse stack of complex
layers --- orders of magnitude fewer unknowns --- to absorb the
multiple scattering outside the VOI, yielding a layer
mathematically equivalent to a confocal transmission image; the second
stage scans this layer and feeds the synthesized angular transmission
fields into a standard BPM-based ODT inversion that
resolves the cellular RI at full voxel resolution.

The framework generalizes across designed phantoms, engineered tissue,
\emph{ex vivo} organs, and \emph{in vivo} bone, whereas standard ODT
applied to the uncorrected reflection signal recovers no meaningful
structure (Fig.~\mbox{\ref{fig:voi-reconstruction}}e). Because RI
reports directly on cellular dry mass and pathological state, mapping it
label-free and in three dimensions deep within intact, opaque tissue ---
without transmission access or destructive sectioning --- enables
applications beyond the reach of conventional ODT: longitudinal
monitoring of opaque organoids, spheroids, and bioprinted constructs for
drug screening and disease modeling; single-cell-level biological
studies deep within living tissue, as with the osteocytes beneath an
intact skull.

Several considerations determine the performance of the proposed
method. First, the method depends on having sufficient intrinsic
backscattering; where the tissue reflectance is too weak, the
optimization can become noise-limited and the reconstruction may fail. A
shorter illumination wavelength would strengthen the backscattering,
albeit at reduced penetration depth. Second, resolution improves as the VOI approaches the reflecting
layer, but the backscattering from the VOI itself can overlap with
round-trip backscattering from the object plane within the temporal
gating window, degrading the fidelity of reconstructions; a
shorter-coherence-length source would relax this trade-off. Third, resolution also depends on the lateral
extent of the recorded field, but enlarging the field of view would slow
imaging --- though at our current field of view the resolution is
already sufficient, the acquisition time of the 1.3\,$\mu$m
\emph{in vivo} system ($\sim$5.9\,s over a \mbox{$120\times120~\mu$m$^{2}$} field
of view) is compatible with \emph{in vivo} imaging. Faster cameras
would ease this speed--area trade-off. Even at its present performance, however, the framework is readily
applicable, extending quantitative phase tomography to opaque
specimens --- biological or industrial --- accessible only in
reflection; addressing the trade-offs above marks an exciting
direction for future work.

\pdfbookmark[1]{Methods}{methods-bm}
\section*{Methods}\label{sec:methods}

\pdfbookmark[2]{Experimental system measuring backscattered waves}{m-exp-bm}
\subsection*{Experimental system measuring backscattered waves}

We built a custom interferometric microscope to record backscattered
waves from biological samples (system schematic in Extended Data Fig.~\ref{fig:ed-system}).
A laser-scanning architecture raster-scans a focused illumination spot
across the sample, and the descanned backscattered field is captured
by cameras placed at planes conjugate to the sample plane. Coherent
detection is implemented in a Mach--Zehnder interferometer driven by a
low-coherence source, so that time-gating selects only photons falling
within the coherence window of the reference beam and rejects
out-of-focus multiply-scattered light.

For high-contrast imaging of cells embedded in scattering tissue, we
used a short visible wavelength generated by second-harmonic generation
(SHG) of a custom-built Yb:KGW femtosecond laser. The mode-locked
fundamental output (center wavelength 1030\,nm, average power 1.8\,W,
pulse duration $\sim$180\,fs, repetition rate 80\,MHz) was frequency-doubled
in a lithium triborate (LBO) crystal, yielding up to 800\,mW at 515\,nm
with a coherence-gating length of ${\sim}50~\mu$m. The beam was focused
through a 60$\times$, NA 1.0 water-dipping objective and steered by
galvanometer mirrors at a step size of 257\,nm and a scan rate of
660 points per second, set by the camera frame rate. With this
configuration, a single volume over a $40\times40~\mu$m$^{2}$ field of
view was acquired in approximately 37\,s. At each scan
position, the descanned sample field was combined with the reference
beam in an off-axis geometry to record an interference hologram, from
which the complex-valued output field was retrieved via Hilbert
transform. A half-wave plate followed by a polarizing beamsplitter set
the sample-to-reference power-splitting ratio, while a linear polarizer
combined with a quarter-wave plate suppressed parasitic back-reflections
from intermediate optical surfaces.

For the \emph{in vivo} through-skull experiment, data were acquired with
a previously established system operating at a 1.3\,$\mu$m center
wavelength but otherwise of similar design to the 515\,nm
setup\cite{kwon2023computational}, whose longer wavelength reduces
scattering and improves penetration through the skull. This system uses a
$\times$25, 1.05\,NA objective (XLPLN25XWMP2, Olympus) and a pulsed source
of 19\,nm bandwidth, yielding a $\sim$25\,$\mu$m time-gating window and a
lateral resolution of $\sim$0.79\,$\mu$m. For this system, the full set
of backscattered waves at a single depth was recorded in ~5.9\,s over an
$120\times120~\mu$m$^{2}$\cite{kwon2023computational}.

\pdfbookmark[2]{Multi-layer fitting of the backscattered waves}{m-fit-bm}
\subsection*{Multi-layer fitting of the backscattered waves}

We approximate the scattering medium as a stack of complex layers at
axial positions \(z_{k}\) measured from the reflecting layer, with
\(z_{0}\equiv 0\) at the reflecting layer and \(z\) increasing toward
the objective (Fig.~\ref{fig:experimental-validation}d). Each layer is
described by its complex amplitude transmittance
\(\Phi_{k}\left( \rho,z_{k} \right)\), except the reflecting layer
\(\Phi_{0}\left( \rho,z_{0} \right)\), which is a complex amplitude
reflectance; \(\rho=(x,y)\) is the lateral coordinate on each layer.
The axial positions \(z_{k}\) were not uniformly spaced. Since the
lateral resolution of a recovered layer \(\Phi_{k}\) decreases with its
distance from the reflecting layer \(z_{0}\), and this lateral
resolution sets the axial resolution attainable at that depth, we
placed the layers more densely near \(z_{0}\) and progressively farther
apart with increasing depth.

For an illumination focused at \(\left( \rho_{\text{in}},z_{0} \right)\)
on the reflecting layer, we model the round trip of the field through
this stack: an incident field \(E_{f}\) enters at the top surface
\(z_{N}\), and a backscattered field \(E_{b}\) returns to the same
plane. All fields carry an implicit dependence on the illumination
position \(\rho_{\text{in}}\), which we suppress below for brevity.

On the downward path, the field after each layer follows the recursion
\begin{equation}
E_{f}\left( \rho,z_{k - 1} \right) = P_{z_{k - 1},z_{k}}\left\{ \Phi_{k}\left( \rho,z_{k} \right)E_{f}\left( \rho,z_{k} \right) \right\},
\label{eq:Ef-recurrence}
\end{equation}
where \(P_{z_{k - 1},z_{k}}\) is the angular-spectrum propagation
kernel,
\begin{equation}
P_{z_{k - 1},z_{k}}\left\{ E \right\} = F^{- 1}\left\{ F\left\{ E \right\}\exp\left( i\left| z_{k} - z_{k - 1} \right|\sqrt{\left( nk_{0} \right)^{2} - \mathbf{q}^{2}} \right) \right\},
\label{eq:asm-kernel}
\end{equation}
with \(\mathbf{q} =\left( k_{x},k_{y} \right)\) the transverse
wavevector, \(k_{0} = 2\pi/\lambda\), and \(n\) the background
RI of the medium. Iterating
Eq.~\eqref{eq:Ef-recurrence} from \(k = N\) down to \(k = 1\) yields the
field \(E_{f}\left( \rho,z_{0} \right)\) at the reflecting layer, which
reflects as
\begin{equation}
E_{b}\left( \rho,z_{0} \right) = \Phi_{0}\left( \rho,z_{0} \right)E_{f}\left( \rho,z_{0} \right).
\label{eq:Eb-reflecting}
\end{equation}
On the upward path, the same propagation is applied in reverse,
\begin{equation}
E_{b}\left( \rho,z_{k} \right) = \Phi_{k}\left( \rho,z_{k} \right)P_{z_{k},z_{k - 1}}\left\{ E_{b}\left( \rho,z_{k - 1} \right) \right\},
\label{eq:Eb-recurrence}
\end{equation}
and iterating Eq.~\eqref{eq:Eb-recurrence} from \(k = 1\) to \(k = N\)
gives the modeled backscattered field \(E_{b}\left( \rho,z_{N} \right)\)
at the surface.

We then optimize the layer set
\(\left\{ \Phi_{k} \mid k = 0,1,\dots,N \right\}\) so that the modeled
field \(E_{b}\) reproduces the measured field \(E_{b}^{\text{GT}}\) at
every illumination, by minimizing
\begin{equation}
\begin{split}
C_{R}\left( \left\{ \Phi_{k} \mid k = 0,1,\dots,N \right\} \right) = {}& -\sum_{\rho_{\text{in}}}\left| \mathrm{corr}_{\rho}\!\left( E_{b}^{\text{GT}},\,E_{b} \right) \right| \\
& + \gamma\sum_{k=0}^{N}\left\| \Phi_{k} \right\|^{2}.
\end{split}
\label{eq:cost-R}
\end{equation}
The first term sums, over all illumination positions
\(\rho_{\text{in}}\), the magnitude of the normalized (Pearson-type)
correlation between the measured and modeled backscattered fields,
\begin{equation*}
\mathrm{corr}_{\rho}(A,B) = \frac{\sum_{\rho}\left( A(\rho) - \overline{A} \right)^{*}\left( B(\rho) - \overline{B} \right)}{\sqrt{\sum_{\rho}\left| A(\rho) - \overline{A} \right|^{2}}\,\sqrt{\sum_{\rho}\left| B(\rho) - \overline{B} \right|^{2}}},
\end{equation*}
where \(\overline{A}\) and \(\overline{B}\) are the spatial means of the
complex fields \(A\) and \(B\); the second term is an \(L_{2}\)
regularization on the layers weighted by \(\gamma\). Because the
backscattered fields in Fig.~\ref{fig:experimental-validation}b were
recorded at a plane conjugate to the reflecting layer, each was
propagated back to the surface \(z_{N}\) before comparison. We
minimized \(C_{R}\) using the ADAM optimizer on the PyTorch platform.

\pdfbookmark[2]{Reconstructing angular transmission fields}{m-atf-bm}
\subsection*{Reconstructing angular transmission fields}

We convert the depth-scanned optimized transmittance layer
\(\Phi_{1}(\rho,z_{1} + m\Delta z)\) into the angular transmission
fields \(E_{p}\left( \rho;\mathbf{q}_{\text{in}},z_{1} \right)\), the
fields transmitted through the volume of interest under plane-wave
illumination at each incident wavevector \(\mathbf{q}_{\text{in}}\).
These fields serve as the inputs to the BPM-based ODT reconstruction. As
described in the main text, each depth-scanned layer is equivalent to the
synthetic-aperture image formed by coherently summing the angular
transmission fields over all illumination angles focused at the
respective depth within the volume of interest. We therefore recover
\(E_{p}\left( \rho;\mathbf{q}_{\text{in}},z_{1} \right)\) by fitting
this synthetic-aperture image to the optimized
\(\Phi_{1}(\rho,z_{1} + m\Delta z)\) at every scanned depth, as
formulated below.

The synthetic aperture image at depth
\(z_{1} + m\Delta z\) can be expressed as a sum over the angular
transmission fields:
\begin{equation}
E_{\text{syn}}\left( \rho,z_{1}+m\Delta z \right) = \sum_{\mathbf{q}_{\text{in}}}^{}{E_{p}\left( \rho;\mathbf{q}_{\text{in}},z_{1}+m\Delta z \right)/E_{\text{bg}}\left( \rho;\mathbf{q}_{\text{in}},z_{1}+m\Delta z \right)}.
\label{eq:syn}
\end{equation}
Here, \(E_{\text{bg}}\left( \rho;\mathbf{q}_{\text{in}},z_{1} \right)\)
is the background field --- the plane-wave illumination field at the
incident wavevector \(\mathbf{q}_{\text{in}}\) in the absence of the
sample --- by which the transmitted field \(E_{p}\) is normalized.
\(E_{p}\left( \rho;\mathbf{q}_{\text{in}},z_{1}+m\Delta z \right)\)
and \(E_{\text{bg}}\left( \rho;\mathbf{q}_{\text{in}},z_{1}+m\Delta z \right)\)
are obtained by free-space propagation of
\(E_{p}\left( \rho;\mathbf{q}_{\text{in}},z_{1} \right)\)
and \(E_{\text{bg}}\left( \rho;\mathbf{q}_{\text{in}},z_{1} \right)\)
over a distance \(m\Delta z\), respectively. We then find the set of
angular transmission fields
\(\left\{ E_{p}\left( \rho;\mathbf{q}_{\text{in}},z_{1} \right) \right\}\)
over all incident wavevectors \(\mathbf{q}_{\text{in}}\) that, across
all scanned depths \(m\), best fits the synthetic-aperture image
\(E_{\text{syn}}\left( \rho,z_{1}+m\Delta z \right)\) to the optimized
transmittance layer \(\Phi_{1}\left( \rho,z_{1}+m\Delta z \right)\), by
minimizing the following cost function:
\begin{equation}
\begin{split}
C_{T}\left( \left\{ E_{p}\left( \rho;\mathbf{q}_{\text{in}},z_{1} \right) \right\} \right) = {}& \sum_{m}\left\| E_{\text{syn}}\left( \rho,z_{1}+m\Delta z \right) - \Phi_{1}\left( \rho,z_{1}+m\Delta z \right) \right\|^{2} \\
& + \tau\sum_{\mathbf{q}_{\text{in}}}\left\| E_{p}\left( \rho;\mathbf{q}_{\text{in}},z_{1} \right) \right\|^{2}.
\end{split}
\label{eq:cost-T}
\end{equation}
Here, the first term sums the fidelity over all scanned depths \(m\),
and \(\tau\) is the hyperparameter that controls the balance between
the fidelity and regularization terms. We minimized \(C_{T}\) using the
ADAM optimizer on the PyTorch platform.

We further note that, in principle, the BPM-based ODT model could be
modified for focused (confocal) illumination and fitted directly to
the optimized transmittance layer \(\Phi_{1}\); however, we found this gave lower
reconstruction fidelity.
Because each layer is a confocal (synthetic-aperture) image whose
amplitude transfer function is weighted toward low spatial
frequencies, fitting directly to \(\Phi_{1}\) under-constrains the
high-frequency components and blurs fine detail. Converting the layers
into angular transmission fields removes this intrinsic low-frequency
bias, so that the standard BPM-based ODT recovers high-frequency
structure faithfully.

\pdfbookmark[2]{Refractive index of the bone matrix and the collagen matrix}{m-bonematrix-bm}
\subsection*{Refractive index of the bone matrix and the collagen matrix}

We determined the average RI $n$ of the bone matrix by
combining two independent measurements on the same skull: the
optical path delay accumulated as light traverses the skull of thickness
$d$, and the apparent focal shift between its top and bottom
surfaces\cite{kang2023sdt}. The optical path delay alone fixes only the
product of $n$ and $d$; the focal shift supplies an independent relation
between the same two quantities, so that solving the pair determines $n$
without prior knowledge of the thickness.

Both measurements were performed with a low-coherence interferometer
(center wavelength 1.3~$\mu$m, coherence length 20~$\mu$m, NA~0.4) on an
excised skull placed on a flat mirror and immersed in PBS
($n_{0} = 1.32$). For the optical path delay, we compared the round-trip
path lengths reflected from mirror regions with and without the skull,
which yields the single-pass delay
\begin{equation}
\Delta z_{\mathrm{ref}} = (n - n_{0})\,d,
\label{eq:opd}
\end{equation}
where $\Delta z_{\mathrm{ref}}$ is taken as half of the measured
round-trip value, since the light double-passes the skull in this
reflection geometry.

For the apparent focal shift, we translated the sample stage and
recorded the axial displacement $\Delta z_{\mathrm{sample}}$ between the
two positions at which the focus coincided with the top and the bottom
surface of the skull. Because the skull has a higher index than the
surrounding medium, the bottom surface appears axially compressed, so
this displacement is smaller than the true thickness; in the paraxial
limit the two are related by
\begin{equation}
d = \frac{n}{n_{0}}\,\Delta z_{\mathrm{sample}},
\label{eq:apparent}
\end{equation}
which is accurate to within 1\% at NA~0.4. Equations~(\ref{eq:opd}) and
(\ref{eq:apparent}) contain only the two measured quantities $\Delta
z_{\mathrm{ref}}$ and $\Delta z_{\mathrm{sample}}$ and the two unknowns
$n$ and $d$. Eliminating $d$ and solving for $n$ gives the refractive
index directly from the measured quantities:
\begin{equation}
n = \frac{n_{0}}{2}\left(1 + \sqrt{\,1 + \frac{4\,\Delta z_{\mathrm{ref}}}{n_{0}\,\Delta z_{\mathrm{sample}}}\,}\,\right).
\end{equation}
This procedure yielded an average RI of the bone matrix of
1.425. The average RI of the collagen matrix (1.339), used as
the background medium index for the cell reconstructions, was determined in
the same manner at the wavelength of 515\,nm.

\pdfbookmark[2]{Quantification of cellular dry mass}{m-drymass-bm}
\subsection*{Quantification of cellular dry mass}

Cellular dry mass was calculated from the RI tomogram by integrating the difference between the local RI $n(\mathbf{r})$ and the solvent RI over the cell
volume,
\begin{equation}
M = \frac{1}{\alpha}\iiint \left[\,n(\mathbf{r}) - n_{\mathrm{solvent}}\,\right]\,dV,
\end{equation}
where $\alpha$ is the specific RI increment and
$n_{\mathrm{solvent}}$ is the RI of the intracellular aqueous solvent
(the cell's water background)\cite{barer1954refractometry}. We took
$n_{\mathrm{solvent}} = 1.32$, the RI of water at the measurement
wavelength $\lambda = 1.3~\mu$m\cite{daimon2007water}. The visible-light
value of $\alpha$ (0.190~mL~g$^{-1}$ at 589~nm)\cite{zhao2011increment}
was extrapolated to 1.3~$\mu$m with the Cauchy dispersion relation\cite{perlmann1948increment},
giving $\alpha \approx 0.181$~mL~g$^{-1}$; this extrapolation beyond the
visible range is approximate.

\pdfbookmark[2]{Animals}{m-animals-bm}
\subsection*{Animals}

Adult C57BL/6 mice (older than 8 weeks) were used in all experiments.
Animals were maintained in a temperature-controlled environment
(20--22\,\textdegree{}C) at a relative humidity of 50--55\% under a
12\,h light / 12\,h dark cycle, with food and water provided
\emph{ad libitum}. All experimental procedures were approved by the
Korea University Institutional Animal Care \& Use Committee
(KUIACUC-2025-0002).

\pdfbookmark[2]{Sample preparation}{m-sampleprep-bm}
\subsection*{Sample preparation}

\pdfbookmark[3]{HT29 cells in 3D collagen matrix}{m-ht29-bm}
\subsubsection*{HT29 cells in 3D collagen matrix}

We prepared 3D engineered tissues made of HT29 human colorectal
adenocarcinoma cells embedded within a fibrous type-I collagen
scaffold. The resulting sample consists of a randomly oriented,
interconnected 3D network of collagen fibers, with
individual fixed cells distributed throughout the inter-fiber pore
space.

\textbf{Cell culture and fixation.} HT29 cells were maintained in
DMEM (LM001-07, Welgene, Gyeongsan-si, Korea) containing 10\% FBS
(S001-01, Welgene) and 1\% penicillin--streptomycin (LS202-02,
Welgene) at 37\,\textdegree{}C in a humidified atmosphere of 5\%
CO\textsubscript{2}. Cells were subcultured at approximately 80\%
confluence by trypsinization, and only cells within passages
32--47 were used. Prior to embedding in the collagen matrix, the cells
were fixed with 4\% paraformaldehyde fixed buffer (420801, BioLegend,
San Diego, USA) for 10\,min at 4\,\textdegree{}C. After fixation,
they were washed three times with fresh PBS (LB 004-02, Welgene) for
5\,min each and stored at 4\,\textdegree{}C.

\textbf{3D collagen matrix and cell embedding.} Two type-I collagen
stocks of different animal origins were used in combination --- one
derived from porcine tendons (Collagen type I gel solution
(3\,mg/mL), (Cellmatrix type 1-A, Nitta Gelatin, Osaka, Japan)) and
one from rat tail (Coll 1 Solution 10\,mg/mL, IKD11926100, CELLINK,
Gothenburg, Sweden). The two stocks were mixed at a volume ratio of
1:1 to yield a final collagen concentration of 6.5~mg/mL; all
mixing steps were carried out on ice to prevent premature gelation.
The cell pellet was added to 200~$\mu$L of 10$\times$ concentrated
MEM-Hanks' medium to achieve a final cell density of
$1.0 \times 10^{6}$\,cells/mL within the gel, and the mixture was
neutralized to physiological pH using 1\,N NaOH.
The cell--collagen mixture was transferred onto a substrate (a bare
glass microscope slide or a gold-coated scratch slide, depending on
the type and size of the experiment) using a pipette and polymerized
at 37\,\textdegree{}C for 1\,hour. After gelation, the cell--collagen
mixture was fixed with 4\% paraformaldehyde fixed buffer for 10\,min
at 4\,\textdegree{}C and washed three times with fresh PBS for 5\,min
each. PBS was then applied to prevent the sample from drying out, and
samples were stored in PBS at 4\,\textdegree{}C until imaging.

\pdfbookmark[3]{In vivo mouse skull preparation}{m-skull-bm}
\subsubsection*{\emph{In vivo} mouse skull preparation}

Mice were anesthetized with isoflurane (1.5--2\% in oxygen;
respiratory rate $\sim$1\,Hz), and body temperature was maintained at
37--38\,\textdegree{}C using a heating blanket. To prevent corneal
drying, ophthalmic ointment was applied to both eyes throughout the
surgical and imaging procedures. After hair removal with a depilatory
cream (Nair), the scalp was carefully removed to expose the parietal
skull, and the remaining connective tissue on the bone surface was
gently cleaned with sterile forceps. A circular glass coverslip
(\#1 thickness, Warner Instruments; 5\,mm diameter, ${\sim}100~\mu$m
thick) was attached directly onto the skull using a UV-curable
adhesive (Loctite 4305). For stable head fixation during imaging, a
custom-made metal head plate was secured to the skull with
cyanoacrylate adhesive, and the surrounding exposed area was reinforced
and sealed with dental cement (Dentsply DeTrey GmbH, Germany). To
minimize postoperative inflammation, dexamethasone (1\,mg/kg) was
administered by intramuscular injection immediately after surgery.
During imaging, anesthesia was maintained with isoflurane (1.2--1.5\%
in oxygen), giving a respiratory rate of $\sim$1.5--2\,Hz, and body
temperature was continuously maintained at 37--38\,\textdegree{}C using
a heating blanket mounted on a three-axis motorized stage.


\pdfbookmark[1]{Acknowledgements}{ack-bm}
\section*{Acknowledgements}

This work was supported by the National Research Foundation of Korea
(NRF) grant No.~RS-2024-00442818 (T.D.H., J.C., T.V.A.N., E.S., J.H.H., S.K., W.C.), No.~RS-2025-24132968 (J.H.H.), No.~RS-2025-24523003
(S.K.), and the Institute of Information \& Communications Technology
Planning \& Evaluation (IITP) grant No.~RS-2025-25464788 (T.V.A.N., E.S.,
S.K., J.H.H., W.C.), funded by the Korea
government (MSIT). This research was also supported by the KERI Primary
research program of MSIT/NST (No.~25A01069) (J.W.K., J.Y.).

\pdfbookmark[1]{Author contributions}{authors-bm}
\section*{Author contributions}

T.D.H. developed the inverse-scattering algorithm including the
multi-layer optimization framework, and carried out the numerical
simulations and tomographic reconstructions.
J.C. designed and built the interferometric microscope together with
E.S. for measuring the time-gated backscattering datasets, and acquired
all of the experimental data on the cell--collagen and mouse kidney
samples.
T.V.A.N. developed the algorithm to construct angular transmission
fields from depth-scanned layers.
J.L. developed the algorithm for tomographic reconstruction from
angular transmission fields.
Y.K. and J.H.H. performed \emph{in vivo} mouse skull imaging.
J.W.K. and J.Y. designed and built the custom Yb:KGW femtosecond
oscillator that served as the second-harmonic illumination source.
J.H.H. prepared all biological samples, including cell--collagen constructs, mouse kidneys, and \emph{in vivo} mouse preparations. 
S.Y., S.K., and W.C. conceived the project. S.Y. and S.K. provided
technical guidance, and S.K. and W.C. supervised the overall project.
T.D.H., J.C., E.S., S.K., and W.C. wrote the manuscript with input from all
authors.

\pdfbookmark[1]{Competing interests}{competing-bm}
\section*{Competing interests}

The authors declare no competing interests.

\clearpage
\pdfbookmark[1]{References}{refs-bm}

\clearpage
\pdfbookmark[1]{Fig. 1: Concept and multi-layer scattering model}{fig-1-bm}

\clearpage
\pdfbookmark[1]{Fig. 2: Multi-layer fitting on a bead/Siemens-star phantom}{fig-2-bm}

\clearpage
\pdfbookmark[1]{Fig. 3: VOI 3D RI reconstruction}{fig-3-bm}

\clearpage
\pdfbookmark[1]{Fig. 4: HT29 cells with intrinsic collagen backscattering}{fig-4-bm}

\clearpage
\pdfbookmark[1]{Fig. 5: In vivo RI tomography within intact mouse skull}{fig-5-bm}

\clearpage
\setcounter{figure}{0}
\renewcommand{\thefigure}{\arabic{figure}}
\renewcommand{\figurename}{Extended Data Fig.}
\pdfbookmark[1]{Extended Data Fig. 1: Schematic of the interferometric microscopy system}{ext-fig-1-bm}
\begin{figure}[htbp]
\centering
\includegraphics[width=0.95\linewidth]{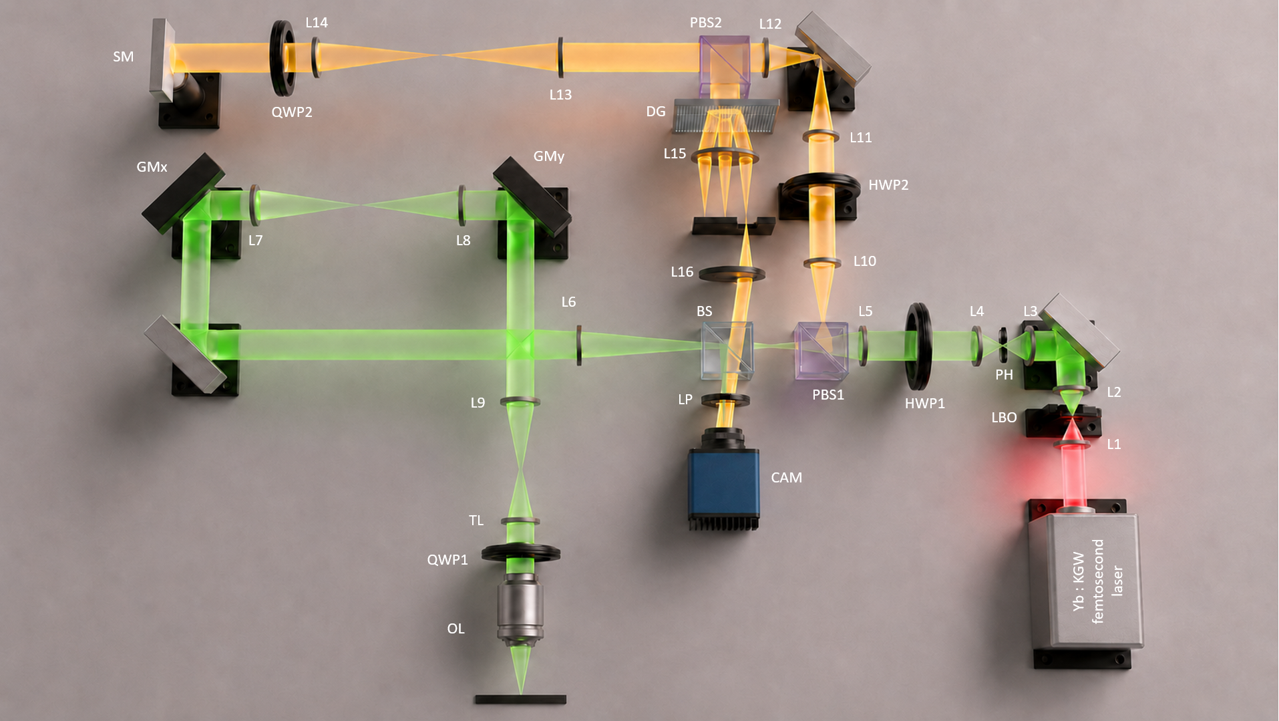}
\caption{\textbf{Schematic of the reflection matrix microscopy system.}
Red, green, and yellow beam paths represent the 1030\,nm fundamental
beam from an Yb:KGW femtosecond laser, the 515\,nm beam illuminating the sample, and the 515\,nm reference
beam, respectively. The system consists of a light source module, a
sample arm, a reference arm, and an off-axis holographic detection
module.
Abbreviations: L1--L16, lenses; HWP1, HWP2, half-wave plates; QWP1,
QWP2, quarter-wave plates; PBS1, PBS2, polarizing beam splitters; BS,
beam splitter; PH, pinhole; GMx, GMy, galvanometer mirrors for the
$x$- and $y$-axis scanning, respectively; TL, tube lens; OL, objective
lens; LP, linear polarizer; DG, diffraction grating; SM, scanning
mirror for coherence-gate adjustment; CAM, camera; LBO, lithium
triborate crystal.}
\label{fig:ed-system}
\end{figure}

\end{document}